\documentclass[aps,pra,twocolumn]{revtex4}
\newcommand {\be}{\begin{equation}}
\newcommand {\ee}{\end{equation}}
\newcommand {\bey}{\begin{eqnarray}}
\newcommand {\eey}{\end{eqnarray}}

\usepackage{epsfig}
\usepackage{amsfonts}
\usepackage{amsmath}
\usepackage{mathbbol}

\usepackage{hyperref}

\begin{document}

\title{Necessary and sufficient optimality conditions for classical simulations 
of quantum communication processes}
\author{Alberto Montina, Stefan Wolf}
\affiliation{Facolt\`a di Informatica, 
Universit\`a della Svizzera Italiana, Via G. Buffi 13, 6900 Lugano, Switzerland}

\date{\today}

\begin{abstract}
We consider the process consisting of preparation, transmission through a 
quantum channel, and subsequent measurement of quantum states. The communication 
complexity of the channel is the minimal amount of classical communication 
required for classically simulating it. Recently, we reduced the computation 
of this quantity to a convex minimization problem with linear constraints. 
Every solution of the constraints provides an upper bound on the communication 
complexity. In this paper, we derive the dual maximization problem of the original one. 
The feasible points of the dual constraints, which are inequalities, give lower bounds on the 
communication complexity, as illustrated with an example. The optimal values of the two 
problems turn out to be equal (zero duality gap). By this property, we provide necessary 
and sufficient conditions for optimality in terms of a set of equalities and inequalities. We 
use these conditions and two reasonable but unproven hypotheses to derive the lower bound 
$n 2^{n-1}$ for a noiseless quantum channel with capacity equal to $n$ qubits. 
This lower bound can have interesting consequences in the context of the recent debate 
on the reality of the quantum state.
\end{abstract}

\maketitle

\section{Introduction}

In some distributed computational tasks, the communication of qubits can replace 
a much larger amount of classical communication~\cite{buhrman}. In some cases,
the gap between classical and quantum communication can even be exponential.
What is the ultimate limit to the power of a quantum channel? In a two-party
scenario,
a limit in terms of classical communication is provided by the communication 
complexity of the channel. As defined in Ref.~\cite{montina}, this quantity
is the minimal amount of classical communication
required to simulate the process of preparation of a state, its transmission 
through the channel and its subsequent measurement. In general, the sender
and receiver can have some restriction on the states and measurements that
can be used. In Ref.~\cite{montina}, we proved that computation of
the communication complexity of a quantum channel is equivalent to a
convex minimization problem with linear constraints, which can be numerically
solved with standard methods~\cite{boyd}. 

In this paper, we derive the dual problem of the previously introduced
optimization problem. A dual problem of a minimization problem (called
primal) is a suitable maximization problem such that the maximum is smaller
than or equal to the minimum of the primal problem. The difference between the 
optimal values of the two problems is called duality gap. If the primal problem
satisfies Slater's condition~\cite{boyd}, then the gap is equal to zero. 
As this condition is satisfied in our case, the new optimization
problem turns out to be equivalent to the original one. We use this property
to show that the primal and dual constraints and an additional equation
are necessary and sufficient conditions for optimality. Points satisfying only
the primal or dual constraints provide upper or lower on the communication
complexity, respectively.

Besides the zero duality gap, the dual reformulation has other interesting
features. First, the number of unknown variables scales linearly in the input 
size. Second, the objective function is linear in the input parameters and the 
variables. Finally, the 
constraints are independent of the input parameters defining the channel.
Thus, if we find the maximum for a particular channel, we can still use
the solution to calculate a lower bound for a different channel, which can
be tight for a slight change of the channel. For example, we could evaluate
the communication complexity for a noiseless quantum channel, and then
find a lower bound for a channel with a weak noise.
We use this reformulation of the original minimization problem
to derive analytically a lower bound for the communication complexity
of a noiseless quantum channel followed by two-outcome projective 
measurements with a rank-$1$ event and its complement. Finally, we use
the necessary and sufficient conditions and two mathematical hypotheses
to derive the lower bound $\frac{1}{2}N\log N$, $N$ being the 
Hilbert~space dimension. Although the hypotheses sound reasonable, we leave them
as an open problem.

The mathematical object under study is an abstract generalization of the 
following physical process. A~sender, Alice, 
prepares a quantum state $|\psi\rangle$. For the moment we assume that she can 
choose the state among a finite set whose elements are labeled by an index $a$.
Second, Alice sends the quantum state to another party, Bob, through a quantum 
channel. Then, Bob performs a measurement chosen among a given set whose
elements are labeled by an index $b$. Again, for the moment we assume
that $b$ takes a finite number of values between $1$ and $M$.
Finally, Bob gets an outcome $s$. 
In a more abstract setting, we consider the overall process as a black
box, which we call C-box, described by a general conditional probability $P(s|a,b)$. 
The C-Box has two inputs $a$ and $b$, which are separately chosen by the
two parties and an outcome $s$, which is obtained by Bob.
This setting goes beyond quantum processes. In particular, it includes the 
communication complexity scenario introduced by 
Yao~\cite{yao}, where $s$ takes two values and $P(s|a,b)$ is deterministic.

A C-box can be simulated classically through a classical channel from Alice
to Bob. We call the minimal communication cost {\it communication
complexity}, denoted by ${\cal C}_{ch}$, of the C-box. Here, we employ an 
entropic definition of communication cost
(see Refs.~\cite{montina,montina2} for a detailed definition). 
Similarly, the asymptotic communication complexity,
denoted by ${\cal C}_{ch}^{asym}$, of a C-box is the minimal asymptotic communication
cost in a parallel simulation of many copies of the C-box.
In Ref.~\cite{montina}, we proved that the asymptotic communication complexity ${\cal C}_{ch}^{asym}$
is the minimum of a convex functional over a suitable space, $\cal V$, of probability
distributions. Then, we also proved a tight lower and upper bound for the
communication complexity ${\cal C}_{ch}$ in terms of ${\cal C}_{ch}^{asym}$.
Namely, we have that,
\be
{\cal C}_{ch}^{asym}\le {\cal C}_{ch}\le{\cal C}_{ch}^{asym}+2\log_2({\cal C}_{ch}^{asym}+1)+2\log_2e.
\ee
Note that a lower bound for the ${\cal C}_{ch}^{asym}$ is also a lower bound
for ${\cal C}_{ch}$. Hereafter, we use the natural logarithm, unless otherwise
specified. Let us define the set $\cal V$.
\newline
{\bf Definition.} Given a C-box $P(s|a;b)$,
the set ${\cal V}$ contains any conditional probability $\rho({\vec s}|a)$
over the sequence ${\vec s}=\{s_1,\dots,s_M\}$
whose marginal distribution of the $b$-th variable is the distribution $P(s|a,b)$ of the
outcome $s$ given $a$ and $b$.
In other words, the set ${\cal V}$ contains any $\rho({\vec s}|a)$ satisfying the
constraints
\be
\label{constraints}
\begin{array}{c}
\rho({\vec s}|a)\ge0,  \\
\sum_{{\vec s},s_b=s} \rho({\vec s}|a)=P(s|a,b),\; \forall  a, b \text{ and } s,
\end{array}
\ee
where the summation is over every component of the sequence $\vec s$, except the
$b$-th component $s_b$, which is set equal to $s$.
\newline
\newline
Then, we proved that
\be
{\cal C}_{ch}^{asym}=\min_{\rho({\vec s}|a)\in{\cal V}} \max_{\rho(a)} {\cal I},
\ee
where 
\be\label{mutual_info}
{\cal I}\equiv 
\sum_{{\vec s},a}\rho({\vec s}|a)
\rho(a)\log\frac{\rho({\vec s}|a)}{\rho(\vec s)}
\ee
is the mutual information between the input and the output~\cite{cover},
and 
\be
\rho(\vec s)=\sum_a \rho(\vec s|a)\rho(a)
\ee
is the marginal distribution of $\vec s$.
As the mutual information is convex in $\rho(\vec s|a)$ and the maximum over a set of 
convex functions is still convex~\cite{boyd}, the asymptotic communication complexity is 
the minimum of a convex function over the space~$\cal V$. Since the set $\cal V$ is
also convex, the minimization problem is convex.

As the mutual information is convex in $\rho({\vec s}|a)$ and concave in $\rho(a)$,
we have from the minimax theorem that
\be
{\cal C}_{ch}^{asym}=\max_{\rho(a)} {\cal J},
\ee
where 
\be\label{F_funct}
{\cal J}\equiv \min_{\rho({\vec s}|a)\in{\cal V}} {\cal I}
\ee
is a functional of the distribution $\rho(a)$.
In some cases, it is trivial to find the distribution $\rho_{max}(a)$ maximizing the
functional $\cal J$.
For example, when there is no restriction on the set of states
and measurements that can be used and the channel is noiseless,
we can infer by symmetry that the distribution $\rho_{max}(a)$ is
uniform. 
Thus, if $\rho_{max}$ is known, the computation of ${\cal C}_{ch}^{asym}$
is reduced to the following convex optimization problem.
\newline
{\bf Problem 1.}
\be\begin{array}{c}
\min_{\rho(\vec s|a)} {\cal I}  \\
\text{subject to the constraints}  \\
\rho({\vec s}|a)\ge0,  \\
\sum_{{\vec s},s_b=s} \rho({\vec s}|a)=P(s|a,b).
\end{array}
\ee
More generally, even if $\rho(a)$ does not maximize the functional $\cal J$, we have 
that ${\cal C}_{ch}^{asym}\ge {\cal J}$.
Thus, the solution of Problem~1 with a non-optimal distribution $\rho(a)$
provides a lower bound on the asymptotic communication complexity. Again, let us recall
that a lower bound for the ${\cal C}_{ch}^{asym}$ is also a lower bound
for ${\cal C}_{ch}$.

\section{Duality}
This section is organized as follows. First, we introduce the concept of 
dual problem of an optimization problem and describe the main properties. 
Then, we derive the dual problem of Problem~1. 
Finally, we show that the primal and dual constraints and an additional
equation are necessary and sufficient conditions for optimality. 
For further details on duality, see Ref.~\cite{boyd}.

\subsection{Dual optimization problem}

Let us consider the following optimization problem:
\be\begin{array}{c}
\min_{\vec x\in D} f(\vec x)  \\
\text{subject to the constraints} \\
g_k(\vec x)=0, \; \forall  k\in\{1,\dots,n\},  \\
h_l(\vec x)\ge0, \; \forall  l\in\{1,\dots,m\},
\end{array}
\ee
where $f(\vec x)$, $g_k(\vec x)$ and $h_l(\vec x)$ are functions 
of a vector $\vec x$ and $D$ is the domain of $f$.

The Lagrangian of this optimization problem is
\be
{\cal L}(\vec x,\vec\lambda,\vec\eta)= f(\vec x)-
\sum_{k=1}^n\lambda_k g_k(\vec x)-\sum_{l=1}^m\eta_l h_l(\vec x),
\ee
where $\vec\lambda$ and $\vec\eta$ are $n$-dimensional and $m$-dimensional
vectors, respectively.

The dual problem is as follows:
\be\begin{array}{c}
\max_{\vec\lambda,\vec\eta} f_{dual}(\vec\lambda,\vec\eta) \\
\text{subject to the constraints} \\
\eta_l\ge0,  \; \forall  l\in\{1,\dots,m\},
\end{array}
\ee
where 
\be
f_{dual}(\vec\lambda,\vec\eta)\equiv\inf_{\vec x\in D} {\cal L}(\vec x,\vec\lambda,\vec\eta).
\ee
For some $\vec\lambda$ and $\vec\eta$, $f_{dual}(\vec\lambda,\vec\eta)$ could be
equal to $-\infty$. This region of the parameters is generally removed by adding
other constraints in the form of a set of inequalities and equalities,
\be\begin{array}{c}
\tilde g_i(\vec\lambda,\vec\eta)=0, \; \forall  i\in\{1,\dots,\tilde n\}. 
\vspace{1mm} \\
\tilde h_j(\vec\lambda,\vec\eta)\ge0, \; \forall  j\in\{1,\dots,\tilde m\}. 
\end{array}
\ee
For points not satisfying these additional constraints, the dual function 
$f_{dual}(\vec\lambda,\vec\eta)$ can be redefined by setting
it equal to any finite value.

Let $p$ and $p_{dual}$ be the optimal values of the primal and dual
problems, respectively. It is easy to check that every feasible point
of the dual problem gives a lower bound on $p$~\cite{boyd}, that is,
\be
\left.
\begin{array}{c}
\eta_l\ge0,  \; \forall  l\in\{1,\dots,m\}  \\
\tilde g_i(\vec\lambda,\vec\eta)=0, \; \forall  i\in\{1,\dots,\tilde n\} 
\vspace{1mm} \\
\tilde h_j(\vec\lambda,\vec\eta)\ge0, \; \forall  j\in\{1,\dots,\tilde m\} 
\end{array}
\right\}
\Rightarrow  f_{dual}(\vec\lambda,\vec\eta)\le p.
\ee
In particular,
\be
\Delta\equiv p-p_{dual}\ge0.
\ee
The difference $\Delta$ between the optimal primal value and the optimal dual
value is called duality gap. When the duality gap is equal to zero, it is said
that strong duality holds. If some mild conditions on the constraints
are satisfied, such as Slater's condition~\cite{boyd}, then $\Delta$ is equal to zero. 
Slater's condition is satisfied if the primal problem is convex and there is a feasible point 
$\vec x_0$ of the constraints such that the strict inequality $h_l(\vec x_0)>0$ holds for 
every $l\in\{1,\dots,m\}$. A refined condition states that strong duality holds if
there is a feasible point $\vec x_0$ such that the strict inequality $h_l(\vec x_0)>0$
holds for every non-affine inequality constraint~\cite{boyd}.
It is easy to realize that Problem~1 satisfies this condition, as every 
constraint is affine and there is at least a feasible point~\cite{montina}.

\subsection{The dual problem of Problem~1}

Let us define the function 
\be\label{geom_object_funct}
{\cal I}_{dual}=\sum_{s,a,b} P(s|a;b)\rho(a)\lambda(s,a,b).
\ee
The dual problem of Problem~1 is as follows.
\newline
{\bf Problem 2.}
\be
\max_{\lambda(s,a,b)} {\cal I}_{dual}
\ee
subject to the constraints
\be\label{constr}
\sum_a\rho(a)e^{\sum_b\lambda(s_b,a,b)}\le 1,\; \forall {\vec s}=(s_1,\dots,s_M).
\ee
The number of variables $\lambda(s,a,b)$ is equal to the number of input parameters 
$P(s|a;b)$, whereas the number of constraints grows exponentially with the number of
measurements. As the primal problem is convex and Slater's refined condition is 
satisfied, strong duality holds and the maximum of ${\cal I}_{dual}$ subject to
constraints~(\ref{constr}) is equal to the solution of Problem~1. 
\newline
{\bf Theorem 1.} Problem~2 is dual to Problem~1 and strong duality holds.
\newline
{\it Proof.}
As already said, strong duality is a direct consequence of the linearity of
the inequality constraints. Let us prove the first statement of the theorem.
The objective function is $\cal I$ defined in Eq.~(\ref{mutual_info}).
The domain $D$ of the objective function is given
by any nonnegative distribution $\rho(\vec s|a)$. 

The Lagrangian of the optimization problem is
\be\label{lagrange}
\begin{array}{c}
{\cal L}={\cal I}- \\
\sum_{s,a,b}\lambda(s,a,b)\rho(a)\left[\sum_{\vec s,s_b=s}\rho(\vec s|a)-P(s|a;b)\right].
\end{array}
\ee
Note that the inequalities $\rho(\vec s|a)\ge0$ are not reckoned as constraints
in the Lagrangian, as they define the domain~$D$ of the objective function.
The Lagrangian can be written in the form
\be\label{L_0_dual}
{\cal L}={\cal L}_0+\sum_{s,a,b}\lambda(s,a,b)\rho(a)P(s|a;b),
\ee
where
\be\label{L_0}
{\cal L}_0\equiv\sum_{\vec s,a}\rho(\vec s|a)\rho(a)
\left[\log\frac{\rho(\vec s|a)}{\rho(\vec s)}-\sum_b\lambda(s_b,a,b)\right]
\ee
The objective function of the dual problem is the infimum of $\cal L$ with
respect to $\rho(\vec s|a)$ in the domain $D$. 
First, let us prove that the infimum is $-\infty$ if
\be\label{dual_c}
\exists \vec s=\vec s\;' \text{ such that }
\sum_a\rho(a)e^{\sum_b\lambda(s_b,a,b)}>1.
\ee
Let us take the distribution
\be
\rho(\vec s|a)=\alpha\delta_{\vec s,\vec s\;'} 
\frac{e^{\sum_b\lambda(s_b,a,b)}}{\sum_{\bar a}\rho(\bar a) 
e^{\sum_{b}\lambda(s_{b},\bar a,b)}},
\ee
where $\alpha$ is any positive real number.
Then,
\be
{\cal L}_0=-\rho(\vec s\;'|a)\rho(a)\log
\sum_{\bar a}\rho(\bar a) e^{\sum_{b}\lambda(s_{b}',\bar a,b)}.
\ee
As $\sum_a\rho(a)e^{\sum_b\lambda(s_b,a,b)}>1$ for $\vec s=\vec s\;'$, we 
have that 
\be
{\cal L}_0<0
\ee
and ${\cal L}_0$ goes to $-\infty$ as $\alpha\rightarrow\infty$.
Thus, we take Ineqs~(\ref{constr}) as constraints of the dual problem in order
to remove this region where $\cal L$ is unbounded below.

Now, let us prove that the infimum of $\cal L$ is the objective function of
Problem~2 when Ineqs.~(\ref{constr}) are satisfied. First, we show that
${\cal L}_0\ge0$. From Eq.~(\ref{L_0}), we have that
\be
{\cal L}_0=-\sum_{\vec s,a}\rho(\vec s|a)\rho(a)\log\left[\frac{\rho(\vec s)}{\rho(\vec s|a)}
e^{\sum_b\lambda(s_b,a,b)}\right].
\ee
As $-\log x$ is a convex function, we have from Jensen's inequality and Ineqs.~(\ref{constr}) 
that
\be
{\cal L}_0\ge-\sum_{\vec s}\rho(\vec s)\log\frac{\sum_{\vec s\;'}\rho(\vec s\;')
\sum_a\rho(a)e^{\sum_b\lambda(s_b',a,b)}}{
\sum_{\vec s\;'}\rho(\vec s\;')}\ge0
\ee
[note that, in general, the distributions $\rho(\vec s|a)\in D$ are not normalized, as
well as $\rho(\vec s)$].
In particular, ${\cal L}_0$ is equal to zero for every $\rho(\vec s|a)$ such
that
\be\label{primal_dual_map}
\rho(\vec s|a)=\rho(\vec s) e^{\sum_b\lambda(s_b,a,b)}.
\ee
Thus, from Eq.~(\ref{L_0_dual}), we have that the infimum of $\cal L$ is
the objective function of Problem~2.
$\square$

By differentiating $\cal L$ with respect to $\rho(\vec s|a)$,
it is easy to show that Eq.~(\ref{primal_dual_map}) is a necessary condition 
for the minimality of the Lagrangian. In particular, 
the solution of Problem~1 must satisfy Eq.~(\ref{primal_dual_map})
for some $\lambda(s,a,b)$.
Note that Eq.~(\ref{primal_dual_map}) implies that
\be
\rho(\vec s)=\rho(\vec s)\sum_a\rho(a)e^{\sum_b\lambda(s_b,a,b)}.
\ee
Thus, $\rho(\vec s|a)$ must be equal to zero for very $a$ if the inequality 
in~(\ref{constr}) is strictly satisfied in $\vec s$.

\subsection{Necessary and sufficient conditions for optimality}

When Eq.~(\ref{primal_dual_map}) is satisfied, the Lagrangian
$\cal L$ turns out to be equal to the dual objective function ${\cal I}_{dual}$. 
If also Ineqs.~(\ref{constr}) are satisfied, then the Lagrangian is smaller 
than or equal to the optimal value $p$, that is,
\be
\left.
\begin{array}{c}
\rho(\vec s|a)=\rho(\vec s) e^{\sum_b\lambda(s_b,a,b)} \\
\sum_a\rho(a) e^{\sum_b\lambda(s_b,a,b)}\le 1.
\end{array}
\right\}\Rightarrow {\cal L}={\cal I}_{dual}\le p.
\ee
If the primal constraints are satisfied, the Lagrangian is 
equal to the primal objective function and it is greater than or equal to the optimal 
value,
\be
\left.
\begin{array}{c}
\sum_{\vec s,s_b=s}\rho(\vec s|a)=P(s|a;b) \\
\rho(\vec s|a)\ge0
\end{array}
\right\}\Rightarrow {\cal L}={\cal I}\ge p.
\ee
Thus, if the primal and dual constraints and Eq.~(\ref{primal_dual_map}) are satisfied,
then the primal and dual objective functions take the optimal value. Note that
the overall constraints could not be simultaneously satisfied if the duality gap
was different from zero. These inferences and the zero duality gap imply the
following.
\newline
{\bf Theorem 2.}
A distribution $\rho(\vec s|a)$ is the solution of Problem~1 if and only if
there is a $\lambda(s,a,b)$ such that the constraints 
\be\label{optimal_cond}
\begin{array}{c}
\rho(\vec s|a)=\rho(\vec s) e^{\sum_b\lambda(s_b,a,b)},   \vspace{1mm} \\
\sum_a\rho(a) e^{\sum_b\lambda(s_b,a,b)}\le 1,  \vspace{1mm}  \\
\sum_{\vec s,s_b=s}\rho(\vec s|a)=P(s|a;b),  \vspace{1mm}  \\
\rho(\vec s|a)\ge0
\end{array}
\ee
are satisfied.

We can replace the last constraint with 
the weaker inequality $\rho(\vec s)\ge0$
because of the first equation and the definition of $\rho(\vec s)$. Using 
the third equation, we can also replace the variables $\rho(\vec s|a)$ in
the first equation with $\rho(\vec s)$. Thus, we get the equivalent conditions
\be
\label{optimal_cond2}
\begin{array}{c}
\rho(\vec s)\ge0, \vspace{1mm} \\
\sum_{\vec s,s_b=s} \rho(\vec s) e^{\sum_{\bar b}\lambda(s_{\bar b},a,\bar b)}
=P(s|a;b),  \vspace{1mm}  \\
\sum_a\rho(a) e^{\sum_b\lambda(s_b,a,b)}\le 1,  \vspace{1mm}  \\
\rho(\vec s)\sum_a\rho(a) e^{\sum_b\lambda(s_b,a,b)}=\rho(\vec s),
\end{array}
\ee
where the last equation ensures that $\sum_a\rho(\vec s|a)\rho(a)=\rho(\vec s)$,
once $\rho(\vec s|a)$ is defined according to the first of Eqs.~(\ref{optimal_cond})
as a function of $\rho(\vec s)$. It is simple to show that the two conditions
are equivalent. Eqs.~(\ref{optimal_cond2}) have the nice property
of reducing the set of unknown variables by replacing $\rho(\vec s|a)$ with
$\rho(\vec s)$.

These conditions turn out to be very useful to check if a numerical or analytic
solution is actually the optimal one. These can also give some hints of the solution,
as we will see in the last section. A simple consequence of these conditions is
the following.\newline
{\bf Corollary 1.}
If 
\be
P(s_0|a_0;b_0)=0 
\ee
for some value of $s_0$, $a_0$ and $b_0$, then one of the solutions of Eqs.~(\ref{optimal_cond})
has
\be
\lambda(s_0|a_0,b_0)=-\infty.
\ee
{\it Proof.} Suppose that there is a solution of the constraints such that
$\lambda(s_0|a_0,b_0)$ is finite. It is easy to realize that Eqs.~(\ref{optimal_cond})
are still satisfied if $\lambda(s_0|a_0,b_0)$ is set equal to $-\infty$.
$\square$

\section{Infinite set of states and measurements}

Until now, we have assumed that Alice and Bob can choose one element in a finite 
set of states and measurements, respectively. In this section, we extend
Problems~1 and 2 to the case of an uncountably infinite number of
states and measurements. Let us consider first the dual Problem~2, which
has an easier generalization.

\subsection{Dual problem}

Let the sets of states and measurements be uncountable and measurable,
the sums over $a$ and $b$ is replaced by integrals. For example,
suppose that Alice can prepare any state and Bob can perform any rank-$1$
projective measurement. Let the dimension of the Hilbert space be $N$. The
space of states is a manifold with dimension $2N-1$ including the physically
irrelevant global phase. The space of measurements is defined as the space
of any orthogonal set of $N$ normalized vectors. Let us denote by 
${\cal M}\equiv(|\phi_1,\dots,|\phi_N\rangle)$ an element in this manifold, where $|\phi_j\rangle$
are the vectors of the orthonormal basis. The function in 
Eq.~(\ref{geom_object_funct}) becomes 
\be\label{obj_continuous}
{\cal I}_{dual}=\sum_s\int d{\cal M}\int d\psi P(s|\psi,{\cal M})\rho(\psi)\lambda(s,\psi,{\cal M})
\ee
in the continuous limit. We choose the integration measure such that
\be
\int d{\cal M}=\int d\psi=1.
\ee
The second equality implies that $\rho(\psi)=1$ if the distribution is uniform
over the space of quantum states. For example, this is the case if there is no
constraint on the set of allowed states and measurements and the channel is noiseless.
Let us denote by $S:{\cal M}\rightarrow s$ any 
function mapping a measurement $\cal M$ to a value $s$ in the set of possible outcomes.
The constraints~(\ref{constr}) become
\be
\int d\psi\rho(\psi) e^{\int d{\cal M} \lambda[S({\cal M}),\psi,{\cal M}]}\le1, \;
\forall\; S .
\ee
These constraints can be recast in the form
\be\label{constr_cont}
\int d\psi\rho(\psi) e^{\sum_s\int_{\Omega_s} d{\cal M} \lambda(s,\psi,{\cal M})}\le1,\;
\forall\; (\Omega_1,\dots,\Omega_N)\in{\cal P},
\ee
where $(\Omega_1,\dots,\Omega_N)\equiv\vec\Omega\in {\cal P}$ is any partition of the 
measurement manifold so that $\Omega_i\cap\Omega_j=\emptyset$ if $i\ne j$ and 
$\cup_i\Omega_i$ is the whole manifold.

Thus, the optimization problem is the maximization of the objective 
function~(\ref{obj_continuous}) under the constraints~(\ref{constr_cont}).
This notation can be generalized to different sets of measurements, such
as operators with degeneration and POVM.

\subsection{Primal problem}
The primal problem~1 can be generalized to the case of the infinite sets of
states $|\psi\rangle$ and measurements $\cal M$ by replacing $\rho(\vec s|a)$ 
with a distribution $\rho(\Omega_1,\dots,\Omega_N|\psi)$ over the
space $\cal P$ of partitions, so that the equality constraint of the
problem is replaced by the equation
\be\label{cont_primal_constr}
\int_{{\cal P}_s({\cal M})} d\Omega\rho(\vec\Omega|\psi)=P(s|\psi,{\cal M}),
\ee
where ${\cal P}_s({\cal M})$ is the set of partitions $(\Omega_1,\dots,\Omega_N)\in{\cal P}$ 
such that ${\cal M}\in\Omega_s$.
The integral in the equation requires the definition of a measure in $\cal P$.
This can be quite problematic. However, as it will be shown in the
next subsection, the optimal distribution
$\rho(\Omega_1,\dots,\Omega_N|\psi)$ is equal to zero for every $|\psi\rangle$
if $\int d\psi\rho(\psi)e^{\sum_s\int_{\Omega_s} d{\cal M} \lambda(s,\psi,{\cal M})}$ is 
strictly smaller than~$1$. Thus, only the subspace of partitions such that 
$\int d\psi\rho(\psi)e^{\sum_s\int_{\Omega_s} d{\cal M} \lambda(s,\psi,{\cal M})}=1$ 
is relevant. This can simplify the definition of the measure, as we need to define
it only in this subspace.

The primal objective function takes the form
\be\label{c_primal_object}
{\cal I}=\int d\psi\int d\Omega\rho(\vec\Omega|\psi)\rho(\psi)\log
\frac{\rho(\vec\Omega|\psi)}{\rho(\vec\Omega)},
\ee
where $\rho(\vec\Omega)=\int d\psi\rho(\vec\Omega|\psi)\rho(\psi)$. The
constraints are Eq.~(\ref{cont_primal_constr}) and the inequalities
$\rho(\vec\Omega|\psi)\ge0$.

\subsection{Necessary and sufficient conditions for optimality}
\label{nec_suff_section}

In the continuous case, the necessary and sufficient conditions~(\ref{optimal_cond})
take the form
\be
\label{c_optimal_cond}
\begin{array}{c}
\rho(\vec\Omega|\psi)=\rho(\vec\Omega) e^{\sum_s\int_{\Omega_s} d{\cal M} \lambda(s,\psi,{\cal M})},
 \vspace{1.5mm} \\
\int d\psi\rho(\psi) e^{\sum_s\int_{\Omega_s} d{\cal M} \lambda(s,\psi,{\cal M})}\le1,
 \vspace{1.5mm}  \\
\int_{{\cal P}_s({\cal M})} d\Omega\rho(\vec\Omega|\psi)=P(s|\psi;{\cal M}),
\vspace{1.5mm}  \\
\rho(\vec\Omega|\psi)\ge0.
\end{array}
\ee
Similarly, the conditions~(\ref{optimal_cond2}) become
\be
\label{c_optimal_cond2}
\begin{array}{c}
\rho(\vec\Omega)\ge0, \vspace{1.5mm} \\
\int_{{\cal P}_s({\cal M})} d\Omega
\rho(\vec\Omega) e^{\sum_{s'}\int_{\Omega_{s'}} d{\cal M}' \lambda(s',\psi,{\cal M}')}
=P(s|\psi;{\cal M}),  \vspace{1.5mm}  \\
\int d\psi\rho(\psi) e^{\sum_s\int_{\Omega_s} d{\cal M} \lambda(s,\psi,{\cal M})}\le 1,  
\vspace{1.5mm}  \\
\rho(\vec\Omega)
\int d\psi\rho(\psi) e^{\sum_s\int_{\Omega_s} d{\cal M} \lambda(s,\psi,{\cal M})}=\rho(\vec\Omega).
\end{array}
\ee
The last equation and the positivity of $\rho(\vec\Omega|\psi)$
imply that $\rho(\vec\Omega|\psi)$ is equal to zero for every $|\psi\rangle$ 
when $\int d\psi\rho(\psi)e^{\sum_s\int_{\Omega_s} d{\cal M} \lambda(s,\psi,{\cal M})}<1$, 
as anticipated in the previous subsection.

We will use these conditions in the last section to argue that the lower bound $n 2^{n-1}$ 
holds for a noiseless quantum channel with capacity equal to $n$ qubits.

\section{Application: lower bounds}
The solution of Problem~2 gives the asymptotic communication complexity of a quantum 
channel. Furthermore, any feasible point satisfying the inequality constraints provides a 
lower bound on ${\cal C}_{ch}^{asym}$ and ${\cal C}_{ch}$. As an application of 
this reformulation of Problem~1,
let us analytically calculate a lower bound in the case of noiseless 
channels and two-outcome measurements with a rank-1 event and its complement. 
The measurement is specified by a vector $|\phi\rangle$ defining the rank-1 event 
$|\phi\rangle\langle\phi|$ and the complement $\mathbb{1}-|\phi\rangle\langle\phi|$.

The objective function and the constraints take the forms (note that $\rho(\psi)=1$)
\be
I=\sum_{s=1}^2\int d{\phi}\int d\psi P(s|\psi,\phi)\lambda(s,\psi,\phi),
\ee
\be\label{constr_2out}
\int d\psi e^{\int_{\Omega} d\phi \lambda(1,\psi,\phi)+
\int_{\Omega^c} d\phi \lambda(2,\psi,\phi)}\le1,\;
\forall\; \Omega,
\ee
where $\Omega$ is a subset of the set of measurements $|\phi\rangle$ and 
$\Omega^c$ is its complement. For a noiseless quantum channel, we have that
\be
P(s|\psi,\phi)=\delta_{s,1}|\langle\psi|\phi\rangle|^2+
\delta_{s,2}(1-|\langle\psi|\phi\rangle|^2).
\ee
The constraints can be written in the form
\be\label{constr_new_form}
\int d\psi e^{\int_{\Omega} d\phi \lambda(\psi,\phi)+
\int d\phi \lambda(2,\psi,\phi)}\le1,\;
\forall\; \Omega,
\ee
where $\lambda(\psi,\phi)\equiv\lambda(1,\psi,\phi)-\lambda(2,\psi,\phi)$.

Every $\lambda(s,\psi,\phi)$ satisfying the constraints induces a lower
bound to the asymptotic communication complexity. Let us consider the simple form 
\be\label{lambda_form}
\lambda(s,\psi,\phi)\equiv \alpha_s |\langle\phi|\psi\rangle|^2+\beta_s
\ee
for these functions.
The constraints are satisfied for a suitable choice of $\alpha_i$ and $\beta_i$.
This is obviously the case for $\alpha_i=\beta_i=0$. 
It is simple to show that
\be
\int d\phi |\langle\phi|\psi\rangle|^2=1/N.
\ee
Furthermore,
\be
\int d\phi |\langle\phi|\psi\rangle|^4=\frac{2}{N(N+1)}.
\ee
Using these equations and defining the variables
$\alpha\equiv\alpha_1-\alpha_2$ and $\beta\equiv\beta_1-\beta_2$,
we have that the objective function takes the form
\be\label{zero_err_funct}
I=\frac{\beta}{N}+\frac{2\alpha}{N(N+1)}+\frac{\alpha_2}{N}+\beta_2
\ee
and the constraints become
\be\label{ineq1}
e^{\frac{\alpha_2}{N}+\beta_2+\beta S_\Omega}\int d\psi e^{\alpha\int_\Omega d\phi 
|\langle\psi|\phi\rangle|^2}\le 1 \;\forall\;\Omega,
\ee
where
\be\label{vol_Omega}
S_\Omega\equiv \int_\Omega d\phi .
\ee

Taking $\Omega$ equal to the empty set and to the whole set of vectors, we get 
the inequalities
\be
\frac{\alpha_2}{N}+\beta_2\le0, \;\;
\frac{\alpha}{N}+\beta+\frac{\alpha_2}{N}+\beta_2\le0.
\ee
To have a non-trivial lower bound, the objective function has to be positive,
thus, the above inequalities and the positivity of $I$ give the following
significant region of parameters
\be\label{signi_region}
\alpha\ge0, \;\;
-\frac{2\alpha}{N+1}\le\beta\le-\frac{\alpha}{N+1}.
\ee
In particular, $\alpha$ must be positive.

Using the Isserlis-Wick theorem~\cite{isserlis}
and the positivity of $\alpha$, it is possible to prove that the left-hand side of
constraint~(\ref{ineq1}) is maximal if $\Omega$ is a suitable cone of vectors. \newline
{\bf Lemma 1.} The left-hand side of the Ineq.~(\ref{ineq1}) is maximal
for a set $\Omega$ such that, for some $|\chi\rangle$ and $\theta\in[0,\pi/2]$,
\be
|\phi\rangle\in\Omega \Longleftrightarrow |\langle\chi|\phi\rangle|^2\ge \cos^2\theta.
\ee
In other words, $\Omega$ is a symmetric cap with symmetry axis $|\chi\rangle$ and angular
aperture $2\theta$.\newline
{\it Proof.}
It is sufficient to prove that the integral in the left-hand side of Ineq.~(\ref{ineq1})
is maximal when $\Omega$ is a symmetric cap for every fixed $S_\Omega$. Using the 
Isserlis-Wick theorem and the positivity of $\alpha$, it is possible to realize that
\be
\int d\psi e^{\alpha\int_\Omega d\phi |\langle\psi|\phi\rangle|^2}=
\sum_{k=0}^\infty c_k \left(\int_\Omega d\phi_1\int_\Omega d\phi_2
|\langle\phi_1|\phi_2\rangle|^2\right)^k,
\ee
where $c_k\ge0$. As $c_k$ are not negative, the integral is maximal for
any fixed $S_\Omega$ when $\int_\Omega d\phi_1\int_\Omega d\phi_2
|\langle\phi_1|\phi_2\rangle|^2$ is maximal. This latter integral is maximal
when $\Omega$ is a symmetric cap. $\square$

Let us denote by $\Omega(\theta)$ a cap with angular aperture~$2\theta$.
From Lemma~1, we have that constraints~(\ref{ineq1}) are satisfied
for any $\Omega$ if and only if they are satisfied for $\Omega=\Omega(\theta)$, where
$\theta$ is any element in $[0,\pi/2]$.
Thus, we need to evaluate the integral in the exponent of the constraints only
over any cone $\Omega(\theta)$ of unit vectors. Let us denote by $S(\theta)$
the quantity $S_{\Omega(\theta)}$. By performing the integral in Eq.~(\ref{vol_Omega}),
we find that
\be
S(\theta)=\sin^{2N-2}\theta.
\ee
Using equation
\be
\int_{\Omega(\theta)} d\phi |\langle\psi|\phi\rangle|^2=S(\theta)\left(\cos^2\theta
|\langle\psi|\chi\rangle|^2
+\frac{\sin^2\theta}{N}\right)
\ee
(see Ref.~\cite{montina2} for its derivation)
and performing the integral over $\psi$ in the constraints~(\ref{ineq1}),
we obtain the inequalities
\be\begin{array}{c}
{\cal F(\theta,\alpha,\beta)}\equiv -S(\theta) \left[ \beta+
\alpha\left(\frac{\sin^2\theta}{N}+\cos^2\theta\right)\right] \\
-\log\frac{\Gamma(N)\left[1-Q(N-1,\alpha S(\theta)\cos^2\theta)\right]}{
\left(\alpha S(\theta)\cos^2\theta\right)^{N-1}}\ge \frac{\alpha_2}{N}+\beta_2,
\;\forall \theta,
\end{array}
\ee
where $Q(x,y)=\Gamma(x,y)/\Gamma(x)$ is the normalized incomplete gamma function, 
$\Gamma(x)$ and $\Gamma(x,y)$ being the complete and incomplete gamma
functions, respectively.

Since the objective function is linear in the unknown variables, at least
one constraint must be active for the optimal values of the parameters.
Thus, the minimum of ${\cal F}(\theta,\alpha,\beta)$ over $\theta$
has to be strictly equal to $\frac{\alpha_2}{N}+\beta_2$.
Let $\theta_m(\alpha,\beta)$ be the value of $\theta$ such that ${\cal F}$ is minimum.
We have that
\bey
{\cal F}[\theta_m(\alpha,\beta),\alpha,\beta]=\frac{\alpha_2}{N}+\beta_2\\
\label{deriv}
\left. \frac{d {\cal F}(\theta,\alpha,\beta)}{d \theta}\right|_{\theta=\theta_m(\alpha,\beta)}=0,
\eey
the last equation coming from the fact that $\theta_m$ is a stationary point in $\theta$.
Note that, until this point, the input function $P(s|\psi,\phi)$ is not involved
in the calculations, as it appears only in the objective function.

Using the first equation, we can remove $\beta_2$ and $\alpha_2$ from
the objective function. We get
\be\label{I_ab}
I=\frac{\beta}{N}+\frac{2\alpha}{N(N+1)}+{\cal F}[\theta_m(\alpha,\beta),\alpha,\beta].
\ee
Now, we assume that the function $\theta_m(\alpha,\beta)$ is differentiable in
the maximal point of $I$. We have checked a posteriori that this turns out to be true for 
$N<5$, but it is false in higher dimensions. The higher dimensional case will be considered
later. Thus, for $N<5$, the objective function is maximal if 
\be
\frac{\partial I}{\partial\alpha}=0, \;
\frac{\partial I}{\partial\beta}=0.
\ee
With Eq.~(\ref{deriv}), we have three equations and three unknown variables,
that is, $\alpha$, $\beta$ and $\theta_m$. To find an analytic solution,
we introduce an approximation by neglecting the normalized incomplete gamma 
function $Q$ in ${\cal F(\theta,\alpha,\beta)}$. Then, we will check the validity of
this approximation. The analytic solution is
\be\label{explicit_alpha}
\alpha=\frac{N^2(N+1)}{N-(N+1)N^\frac{1}{1-N}}, 
\ee
\be\label{cond_theta}
S(\theta_m)=\sin^{2N-2}\theta_m=\frac{1}{N}, 
\ee
\be\label{sol_beta}
\beta=\left(\frac{\left(1-N^\frac{1}{1-N}\right)^{-1}-1}{N}-2\right)\frac{\alpha}{N+1}.
\ee

Using these equations, we obtain that the maximum is
\be\label{maxim}
I_{max}=(N-1)\log\frac{N(N+1)\left(N^\frac{1}{1-N}-1\right)e^{-1}}{
\left[(1+N)N^\frac{1}{1-N}-N\right]\Gamma^\frac{1}{N-1}(N)}.
\ee
Thus, in base $2$ of the logarithm, we have the lower bounds
$1.14227$, $1.86776$, and $2.45238$ bits for $N=2,3,4$, respectively.
They are higher than the trivial lower bound of $1$ bit, which is
the classical information that can be communicated through the
channel with subsequent two-outcome measurement. They even beat
the trivial bounds obtained in the case of rank-$1$ measurements, 
$\log_22=1$, $\log_23=1.585$, and $\log_24=2$, although we considered
only simulations of a channel with subsequent two-outcome measurements.
\begin{figure}
\epsfig{figure=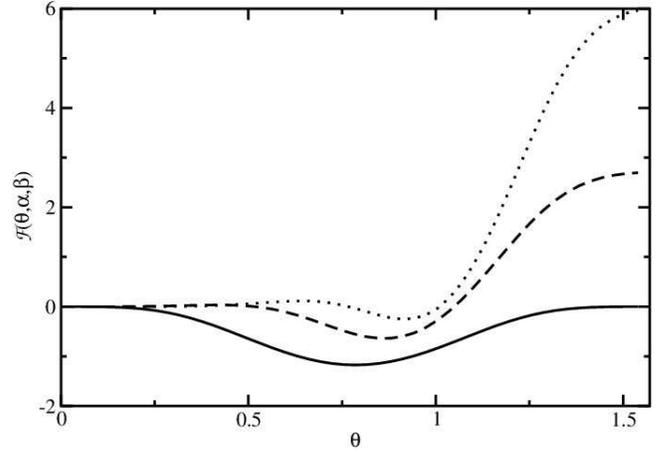,width=8.5cm}
%\vspace{1mm}
\caption{${\cal F}(\theta,\alpha,\beta)$ as a function of $\theta$ for
$N=2$ (solid line), $N=3$ (dashed line) and $N=4$ (dotted line). The
variables $\alpha$ and $\beta$ take the values maximizing the objective
function $I$ in Eq.~(\ref{I_ab}). The minimum of $\cal F$ is in 
$\theta=\arcsin N^\frac{1}{2-2N}$, in agreement with Eq.~(\ref{cond_theta}). }
\label{fig1}
\end{figure}

To derive Eq.~(\ref{maxim}), we have neglected the normalized incomplete gamma 
function $Q$ in ${\cal F}(\theta,\alpha,\beta)$. The exact solution still satisfies
Eqs.~(\ref{cond_theta},\ref{sol_beta}), but the explicit Eq.~(\ref{explicit_alpha})
is replaced by the implicit equation for $\alpha$
\be\label{exact_eq_alpha}
\begin{array}{c}
\left(N^\frac{N}{1-N}-\frac{1}{N+1}\right)\frac{\alpha}{N}+1= \vspace{1mm} 
\frac{e^{-\frac{\cos^2\theta_m\alpha}{N}} \left(\frac{\cos^2\theta_m\alpha}{N}\right)^{N-1}
}{\Gamma(N)\left[1-Q(N-1,\cos\theta_m\alpha/N)\right]},
\end{array}
\ee
where $\theta_m$ is given by Eq.~(\ref{cond_theta}). The approximate $\alpha$
given by Eq.~(\ref{explicit_alpha}) is obtained by neglecting the right-side
term in Eq.~(\ref{exact_eq_alpha}).

To check the validity of the approximation used to calculate the maximum~(\ref{maxim}),
we have numerically solved the exact Eq.~(\ref{exact_eq_alpha}) through few iterations
of the Newton method. We obtain slightly higher values, thus
Eq.~(\ref{maxim}) gives an exact valid lower bound. The numerical bounds
are $1.14602$, $1.87606$, and $2.46463$ bits for $N=2,3,4$, respectively.
Note that Eq.~(\ref{deriv}) guarantees that $\theta_m$ is a stationary
point of ${\cal F}$, not a minimum. To be sure that $\theta_m$ is actually
a minimum, we have evaluated ${\cal F}$ as a function of $\theta$, see
Fig.~\ref{fig1}.

\begin{figure}
\epsfig{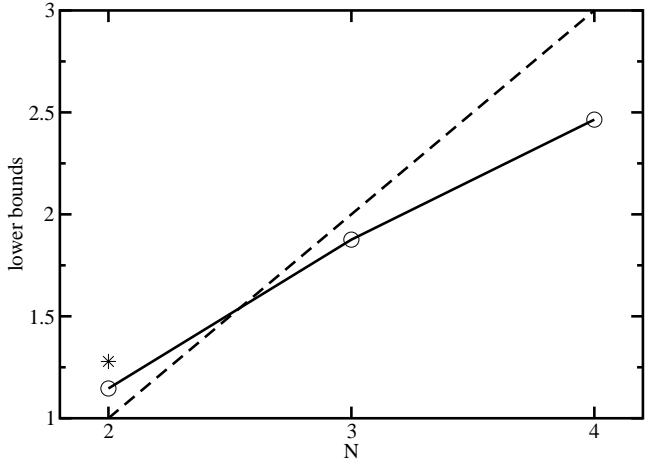}
%\vspace{1mm}
\caption{Calculated lower bound of the communication cost (solid line). The dashed line 
is the lower bound proved using the double cap conjecture~\cite{montina3}. The
star for $N=2$ is the lower bound obtained in Ref.~\cite{montina}. As the measurements
have two outcomes and the quantum channel is noiseless, $1$ bit is a trivial lower bound.}
\label{fig2}
\end{figure}

The lower bound for $N=2$ is lower than the bound $1+\log_2\frac{\pi}{e}\simeq1.2088$
previously derived in Ref.~\cite{montina}. Also, the other two bounds are lower 
than the bound $N-1$ proved in Ref.~\cite{montina3}, but the proof
relies on an unproven property, called double-cap conjecture. The overall
bounds are plotted in Fig.~\ref{fig2}. If we extrapolated Eq.~(\ref{maxim}),
we would have that the lower bound for high $N$ would scale as
\be
I_{max}\sim N \log\left(1+\frac{1}{\log N}\right)\sim N/\log N,
\ee
which is sublinear in $N$. However this asymptotic behavior is not reliable, as
Eq.~(\ref{maxim}) does not hold for $N>4$.

Let us consider the case $N>4$. In this case, we have noted that $\theta_m(\alpha,\beta)$
is not differentiable in the maximal point. This comes from the fact that the 
function ${\cal F}(\theta,\alpha,\beta)$ has two global minima for the optimal values
of $\alpha$ and $\beta$. One minimum is at $\theta=0$, the other one is at some
$\theta_m$ different form zero. The minimum of ${\cal F}$ turns out to be equal to
zero, that is,
\be
{\cal F}(0,\alpha,\beta)={\cal F}(\theta_m,\alpha,\beta)=0
\ee
for optimal $\alpha$ and $\beta$.
When $\alpha$ and $\beta$ are changed to some
suitable direction, the second minimum becomes smaller than the
first one. The opposite occurs if we move to the opposite direction. This
gives the discontinuity of $\theta_m(\alpha,\beta)$.

The optimal solution is found by solving the equations
\be\label{optimalN5}
\begin{array}{c}
{\cal F}(\theta_m,\alpha,\beta)=0, \vspace{1.5mm} \\
\left.\frac{d {\cal F}(\theta,\alpha,\beta)}{d \theta}\right|_{\theta=\theta_m}=0, 
\vspace{1.5mm} \\
\frac{\partial{\cal F}}{\partial\alpha}\frac{\partial I}{\partial\beta} -
\frac{\partial{\cal F}}{\partial\beta}\frac{\partial I}{\partial\alpha}=0.
\end{array}
\ee
The first two equations are implied by the condition that $\cal F$ is equal to 
zero at a stationary point $\theta=\theta_m$. The last equation requires 
that the objective function is stationary for perturbations of $\alpha$ and 
$\beta$ that do not change the minimum of $\cal F$.
Denoting by $S$ and $\tau$ the values $S(\theta_m)$ and $\cos(\theta_m)S(\theta_m)$, 
respectively, the last equation gives the equality
\be\label{last_eq}
\frac{e^{\tau\alpha}(\tau\alpha)^{N-1}}{\Gamma(N)[1-Q(N-1,\tau\alpha)]}=
1-\left(\frac{1}{1+N}-\frac{\sin^2\theta_m}{N}\right)\alpha S,
\ee
The last and second of Eqs.~(\ref{optimalN5}) give the equation
\be\label{beta}
\beta=\alpha\frac{\tan^2\theta_m-2N}{N(N+1)}.
\ee
Eqs.~(\ref{last_eq},\ref{beta}) and the first of Eqs.~(\ref{optimalN5}) give
the explicit equation for $\alpha$
\be
\alpha=-\frac{N(N+1)\left[g+W_0(-g e^{-g})\right]}{S(1+\cos^2\theta_m)
(\tan^2\theta_m-N)},
\ee
where $g\equiv 1+\cos^{-2}\theta_m$ and $W_0(x)$ is the Lambert $W$ function.
Using this equation in Eq.~(\ref{last_eq}) to eliminate $\alpha$, we get an
equation for $\theta_m$, which we can numerically solve with the additional
condition that $\theta=\theta_m$ is a global minimum of 
${\cal F}(\theta,\alpha,\beta)$.

\begin{figure}
\epsfig{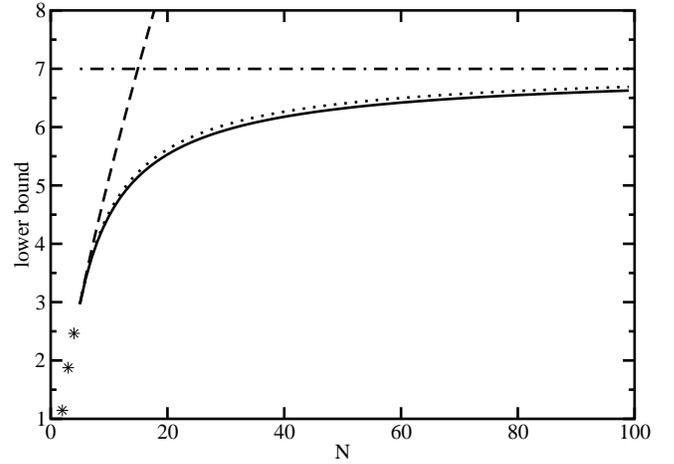}
%\vspace{1mm}
\caption{The solid line is the numerically evaluated lower bound on the
communication complexity for $N\ge5$. The dotted line is the analytic
asymptotic limit~(\ref{asympt}). The lower bound converges to the value
indicated by dash-dotted line in the limit $N\rightarrow\infty$. The dashed
line is the extrapolation of Eq.~(\ref{maxim}) above $N=4$. The stars
are the lower bounds given by Eq.~(\ref{maxim}) for $N\le4$. }
\label{fig3}
\end{figure}

The computed lower bound for $N$ between $5$ and $100$ is plotted in Fig.~\ref{fig3}
(solid line). We have double-checked the calculations by computing the
lower bound for low values of $N$ with a brute-force Monte Carlo method and 
the aid of Ineqs.~(\ref{signi_region}).
For the sake of comparison, we also report the extrapolation of 
Eq.~(\ref{maxim}) above $N=4$ (dashed line).
It is possible to show that the lower bound of the
communication complexity asymptotically converges to
\be\label{asympt}
\frac{[N(e-z_1)+z_1]^2}{N z_1(N e+z_1) \left(1+\frac{z_1}{N(e-z_1)}\right)^N  }.
\ee
for $N\rightarrow\infty$ (in the natural base of the logarithm),
where $z_1$ is the second solution of the equation
\be
1-\frac{e}{z}-\frac{z}{e}+\log z=0
\ee
and it is about $6.895$. 
This asymptotic limit is a very good approximation of the lower bound for
every $N\ge5$, as shown in Fig.~\ref{fig3} (dotted line).
In the limit $N\rightarrow\infty$, the lower bound
saturates to $e^{\frac{e}{z_1-e}}\frac{(z_1-e)^2}{z_1\log 2}$ in base $2$, which 
is equal to about $6.998$ bits (dash-dotted line in Fig.~\ref{fig3}). Thus, the 
found lower bound is very loose in high dimension.

\section{Conjecturing a better lower bound}

In the previous section, we have evaluated a lower bound on the communication
complexity of a quantum channel in the case of projective measurements with
a rank-$1$ event and its complement. This has been achieved by considering
a subset of functions $\lambda(s,\psi,\phi)$ parametrized by a suitable 
finite set of variables [see Eq.~(\ref{lambda_form})].
We have found that the set $\Omega(\theta_m)$ changes 
discontinuously with respect to the parameters in the optimal point for $N>4$. 
Here, we assume that this discontinuity disappears around the optimal
solution when the full space of $\lambda$'s is considered. Using this hypothesis and another
assumption, we are able to prove the lower bound $\frac{1}{2}N\log N=n2^{n-1}$,
$n$ qubits being the quantum capacity of the channel. For this purpose, 
the necessary and sufficient conditions in Sec.~\ref{nec_suff_section} are used.

Let $\lambda_0(s,\psi,\phi)$ be the solution of the optimization Problem~2
for the quantum communication process considered in the previous section.
We define the function
\be
G(\Omega)\equiv\int d\psi e^{\int_{\Omega} d\phi \lambda_0(\psi,\phi)+
\int d\phi \lambda_0(2,\psi,\phi)},
\ee
where $\lambda_0(\psi,\phi)\equiv\lambda_0(1,\psi,\phi)-\lambda_0(2,\psi,\phi)$.
We have from Ineq.~(\ref{constr_new_form}) [corresponding to the second of
Eqs.~(\ref{c_optimal_cond})] that 
\be\label{dual_c2}
G(\Omega)\le 1, \; \forall \Omega.
\ee
Let $\bar\Omega_m$ be any set maximizing the left-hand side of this inequality. 
As noted in the previous section, since the objective function is
linear in $\lambda(s,\psi,\phi)$, at least some inequality constraints
must be active. Thus, we necessarily have that
\be
G(\bar\Omega_m)=1.
\ee
Now, we introduce a perturbation to the solution by taking
\be
\lambda(s,\psi,\phi)=\lambda_0(s,\psi,\phi)+\beta_s,
\ee
where $\beta_1$ and $\beta_2$ are real parameters such that the inequality
constraints of the optimization problem still hold.
Ineq.~(\ref{constr_new_form}) becomes
\be
G(\Omega)e^{\beta S_\Omega+\beta_2}\le1,
\ee
where $\beta\equiv\beta_1-\beta_2$. Let $\Omega_m(\beta,\beta_2)$ be a set 
$\Omega$ maximizing the left-hand side of the inequality. We take $\beta_2$
such that the constraint is active for $\Omega=\Omega_m(\beta,\beta_2)$, that
is,
\be\label{constr_pert}
G(\Omega_m(\beta,\beta_2))e^{\beta S_{\Omega_m(\beta,\beta_2)}+\beta_2}=1.
\ee
For $\beta=\beta_2=0$, $\Omega_m(\beta,\beta_2)$ is the set $\bar\Omega_m$.
The objective function is
\be
{\cal I}_{dual}=I_m+\frac{\beta}{N}+\beta_2,
\ee
where $I_m$ is the maximum, achieved when $\beta=\beta_2=0$.
Using Eq.~(\ref{constr_pert}), we can write the objective function as
\be\label{dual_transf}
{\cal I}_{dual}=I_m+\beta\left[\frac{1}{N}-S_{\Omega_m(\beta,\beta_2)}\right]-
\log[G(\Omega_m(\beta,\beta_2))].
\ee
Assuming that the objective function is differentiable at the
maximum, we have that
\be\label{min_condition}
\left.\frac{{\partial\cal I}_{dual}}{\partial\beta}\right|_{\beta=\beta_2=0}=0.
\ee
Furthermore, since $G$ is maximal for $\Omega=\bar\Omega_m$, we assume
that $G$ is stationary for infinitesimal $\beta$. 
Thus,
\be\label{conjecture}
\left.\frac{\partial G(\Omega_m(\beta,\beta_2))}{\partial\beta}\right|_{\beta=\beta_2=0}=0.
\ee
This equation and Eq.~(\ref{min_condition}) give the condition
\be
S_{\bar\Omega_m}=\frac{1}{N},
\ee
which is identical to Eq.~(\ref{cond_theta}) in the previous section.
Thus, the sets $\Omega$ maximizing the left-hand side of the inequality 
constraints~(\ref{dual_c2}) has a volume equal to $1/N$. For these
sets, the constraint is active, that is,
\be
\int d\psi e^{\int_{\bar\Omega_m} d\phi \lambda(1,\psi,\phi)+
\int_{\bar\Omega_m^c} d\phi \lambda(2,\psi,\phi)}=1,
\ee
where $\bar\Omega_m^c$ is the complement of $\bar\Omega_m$.

From the last of Eqs.~(\ref{c_optimal_cond2}) and the positivity of $\rho(\vec s|\psi)$
[last of Eqs.~(\ref{c_optimal_cond})], 
we have that $\rho(\Omega,\Omega_c|\psi)$
is equal to zero for every $|\psi\rangle$ when $S_\Omega\ne\frac{1}{N}$. From the third and
the last of Eqs.~(\ref{c_optimal_cond}), we have that $\rho(\Omega,\Omega_c|\psi)$ is
equal to zero when $|\psi\rangle$ is orthogonal to some vector $|\phi\rangle$ 
in $\Omega$. Let us recall that $|\phi\rangle$ defines the two-outcome projective 
measurement with event $|\phi\rangle\langle\phi|$ and its complement.
Thus,
\be
\rho(\Omega,\Omega_c|\psi)\ne0\Rightarrow \left\{
\begin{array}{c}
\Omega=\bar\Omega_m\Rightarrow S_\Omega=\frac{1}{N} \vspace{1.5mm} \\
\forall |\phi\rangle\in\Omega, \; \langle\phi|\psi\rangle\ne0
\end{array}\right. .
\ee
Now, we assume that the claim of Lemma~1 still holds, that is, we assume
that the maximal sets
$\bar\Omega_m$ are symmetric caps. Their angular
aperture $2\theta$ is such that $S(\theta)=\sin^{2N-2}\theta=\frac{1}{N}$. It is 
easy to realize that the maximal set of vectors $|\psi\rangle$ which are not 
orthogonal to every vector in $\Omega=\bar\Omega_m$ is a symmetric cap with angular 
aperture $\pi-2\theta$ and same symmetry axis of $\bar\Omega_m$. Its volume is 
$S(\pi/2-\theta)=\cos^{2N-2}\theta=\left(1-N^{\frac{1}{1-N}} \right)^{N-1}\equiv 
S_{max}$. Thus, for every partition $(\Omega,\Omega^c)$, 
$\rho(\Omega,\Omega^c|\psi)$ is different from zero in a set of $|\psi\rangle$ whose
volume is not greater than $S_{max}$. Since the distribution $\rho(\psi)$ is
uniform and different from zero for every $|\psi\rangle$, the mutual information between
the partition $(\Omega,\Omega^c)$ and $|\psi\rangle$ is equal to or greater than
$-\log_2 S_{\max}$. This can be shown by writing the mutual information
in Eq.~(\ref{c_primal_object}) in the form
\be
{\cal I}=\int d\psi\int d\Omega\rho(\psi|\vec\Omega)\rho(\vec\Omega)\log
\frac{\rho(\psi|\vec\Omega)}{\rho(\psi)}
\ee
(note that $\rho(\psi)=1$).
Thus,
\be
{\cal C}_{ch}^{asym}\ge -(N-1)\log \left(1-N^{\frac{1}{1-N}} \right),
\ee
which implies the inequality
\be
{\cal C}_{ch}^{asym}\ge \frac{1}{2} N\log N= n 2^{n-1}.
\ee
This inequality holds for every $N$, but in high dimension the factor $1/2$ can
be replaced by a higher number approaching $1$ as $N\rightarrow\infty$.

Our derivation relies on Eq.~(\ref{conjecture}) and the hypothesis that the maximal
sets $\bar\Omega_m$ are symmetric caps. In particular, the first property could be
false. However, these two hypotheses can be used to simplify greatly the analytic 
derivation of the explicit solution of the optimization problem. Then, we can use 
the necessary and sufficient conditions to check the solution 
and, thus, the validity of the hypotheses. This strategy will be object of future 
investigation.

\section{Conclusion}

In Ref.~\cite{montina}, we reduced the computation of the communication complexity
of a quantum channel to a convex minimization problem with equality constraints. 
Here, we have derived the dual maximization problem. The dual constraints
are given by inequalities. Feasible points of the 
constraints provide lower bounds on the communication cost. We have used this 
optimization problem to derive analytically some non-trivial lower bounds for a 
noiseless quantum channel and subsequent two-outcome measurements with a rank-$1$ 
event and its complement. Furthermore, we have provided necessary and sufficient
conditions for optimality in terms of a set of equalities and inequalities.
Using these conditions and two additional hypotheses, we have derived the
lower bound $\frac{1}{2}N\log N$. Although the two hypotheses sound reasonable,
they have to be proved. One strategy is to use these hypotheses to simplify 
considerably the derivation of the explicit solution of the optimization problem. 
Once the solution is found, we can use the necessary and sufficient conditions to
check the validity of the hypotheses. This route will be undertaken in future
works. The lower bound $\frac{1}{2}N\log N$ would have interesting
consequences in the context of the recent debate on the reality of the quantum
state~\cite{pusey,montina5}. Indeed, this would imply that the support
of the distribution of quantum states given the hidden-variable state
would collapse (in a sufficiently fast way) to a zero-measure set in 
the limit of infinite qubits.
The relation between this quantum foundational problem and 
the communication complexity of a quantum channel was already 
pointed out in Ref.~\cite{montina5}.

{\it Acknowledgments.} 
This work is supported by the Swiss National Science Foundation, 
the NCCR QSIT, and the COST action on Fundamental Problems in Quantum Physics.

\bibliography{biblio.bib}

\begin{thebibliography}{20}



\bibitem{buhrman} H. Buhrman, R. Cleve, S. Massar, and R. de Wolf, Rev. Mod. Phys. {\bf 82}, 665 (2010).
\bibitem{montina} A. Montina, M. Pfaffhauser, S. Wolf, Phys. Rev. Lett. {\bf 111}, 160502 (2013).
\bibitem{boyd} S. Boyd, L. Vandenberghe, {\it Convex Optimization} 
(Cambridge University Press, Cambridge, 2004).
\bibitem{cover} T. M. Cover and J. A. Thomas, {\it Elements of Information Theory} (Wiley, New York, 1991).
\bibitem{yao} A. C. Yao, Proc. of 11th STOC {\bf 14}, 209 (1979).
\bibitem{montina2} A. Montina, Phys. Rev. A {\bf 87}, 042331 (2013). 
\bibitem{isserlis} L. Isserlis,  Biometrika {\bf 11} 185 (1916);
Wick, G.C. (1950), Phys. Rev. {\bf 80} 268 (1950).
\bibitem{montina3} A. Montina, Phys. Rev. A {\bf 84}, 060303(R) (2011).
\bibitem{pusey} M. F. Pusey, J. Barrett, T. Rudolph, Nature Physics {\bf 8}, 476 (2012);
R. Colbeck, R. Renner, Phys. Rev. Lett. {\bf 108}, 150402 (2012);
M. Schlosshauer, A. Fine, Phys. Rev. Lett. {\bf 108}, 260404 (2012);
L. Hardy, arXiv:1205.1439.
\bibitem{montina5} A. Montina, Phys. Rev. Lett. {\bf 109}, 110501 (2012).
%\bibitem{toner} B. F. Toner and D. Bacon, Phys. Rev. Lett. {\bf 91}, 187904 (2003).
%\bibitem{montina} A. Montina, Phys. Rev. Lett. {\bf 109}, 110501 (2012).
%\bibitem{montina2} A. Montina, Phys. Rev. A {\bf 84}, 060303(R) (2011).
%\bibitem{rev_s} C. H. Bennett, P. Shor, J. Smolin, and A. V. Thapliyal, 
%IEEE Trans. Inf. Theory, {\bf 48} 2637 (2002).
%\bibitem{montina3} A. Montina, Phys. Rev. A {\bf 87}, 042331 (2013). 
%\bibitem{wilde} M. Berta, J. M. Renes, and M. M. Wilde, arXiv:1301.1594; M. M. Wilde, P. Hayden, 
%F. Buscemi, M-H. Hsieh, arXiv:1206.4121.
%\bibitem{cover} T. M. Cover and J. A. Thomas, {\it Elements of Information Theory} (Wiley, New York, 1991).
%\bibitem{harsha} P. Harsha, R. Jain, D. McAllester, J. Radhakrishnan,
%IEEE Trans. Inf. Theory {\bf 56}, 438 (2010).
%\bibitem{ks} S. Kochen and E. Specker, J. Math. Mech. {\bf 17}, 59 (1967).

\end{thebibliography}

\end{document}